\documentclass[aps, pra, reprint,a4paper, amsfonts, amssymb, amsmath, reprint, showkeys, nofootinbib,superscriptaddress]{revtex4-1}
\usepackage[english]{babel}
\usepackage[utf8]{inputenc}
\usepackage{float}
\usepackage[colorinlistoftodos, color=green!40, prependcaption]{todonotes}
\usepackage{soul,xcolor}
\setstcolor{red}

\usepackage{amsthm}
\usepackage{mathtools}
\usepackage{physics}
\usepackage{xcolor}
\usepackage{graphicx}
\usepackage[left=20mm,right=20mm,top=35mm,columnsep=15pt]{geometry} 
\usepackage{adjustbox}
\usepackage{placeins}
\usepackage[T1]{fontenc}
\usepackage{lipsum}
\usepackage{csquotes}
\usepackage{wrapfig}
\usepackage{placeins}
\usepackage{appendix}
\usepackage[pdftex, pdftitle={Article}, pdfauthor={Author}]{hyperref} 

\hypersetup{
    colorlinks = true,
    allcolors = {blue},
}

\newcommand*{\refApp}[1]{
    \textcolor{blue}{Appendix} \ref{#1}
}

\makeatletter
\renewcommand*{\@fnsymbol}[1]{\ensuremath{\ifcase#1\or 
   \mathsection\or \mathparagraph\or \|\or **\or \dagger\dagger
   \or \ddagger\ddagger \else\@ctrerr\fi}}
\makeatother

\begin{document}
\title{Tuneable spin-glass optical simulator based on multiple light scattering}
\author{Gianni Jacucci}
\thanks{These two authors contributed equally}
\author{Louis Delloye}
\thanks{These two authors contributed equally}
\affiliation{Laboratoire Kastler Brossel, Sorbonne Université, Ecole Normale Supérieure-Paris Sciences et Lettres (PSL) Research University, Centre Nationale de la Recherche Scientifique (CNRS) UMR 8552, Collège de France, 24 rue Lhomond, 75005 Paris, France}

\author{Davide Pierangeli}
\affiliation{Institute for Complex System, National Research Council (ISC-CNR), 00185 Rome, Italy}
\affiliation{Dipartimento di Fisica, Università di Roma “La Sapienza”, 00185 Rome, Italy}

\author{Mushegh Rafayelyan}
\affiliation{Department of Physics, Yerevan State University, 0025 Yerevan, Armenia}

\author{Claudio Conti}
\affiliation{Institute for Complex System, National Research Council (ISC-CNR), 00185 Rome, Italy}
\affiliation{Dipartimento di Fisica, Università di Roma “La Sapienza”, 00185 Rome, Italy}

\author{Sylvain Gigan}
\affiliation{Laboratoire Kastler Brossel, Sorbonne Université, Ecole Normale Supérieure-Paris Sciences et Lettres (PSL) Research University, Centre Nationale de la Recherche Scientifique (CNRS) UMR 8552, Collège de France, 24 rue Lhomond, 75005 Paris, France}

\begin{abstract}
The race to heuristically solve non-deterministic polynomial-time (NP) problems through efficient methods is ongoing. 
Recently, optics was demonstrated as a promising tool to find the ground-state of a spin-glass Ising Hamiltonian, which represents an archetypal NP problem.
However, achieving completely programmable spin couplings in these large-scale optical Ising simulators remains an open challenge.
Here, by exploiting the knowledge of the transmission matrix of a random medium, we experimentally demonstrate the possibility of controlling the couplings of a fully-connected Ising spin-system.
By tailoring the input wavefront we showcase the possibility of modifying the Ising Hamiltonian both by accounting for an external magnetic-field and by controlling the number of degenerate ground-states and their properties and probabilities. 
Our results represent a relevant step toward the realisation of fully-programmable Ising-machines on thin optical-platforms, capable of solving complex spin-glass Hamiltonians on a large scale.
\end{abstract}

\maketitle
\section{Introduction} \label{sec:intro}

Non-deterministic polynomial-time problems, also known as NP-problems, are tasks whose solve-time scales exponentially with the size of the input \cite{np-pbl}. Such problems appear in most domains whether it is economy, society, science, and cryptography \cite{np-completeness}. 
Importantly, \textcolor{black}{NP-complete problems can be mapped onto each other, solving one solves all}. Among these, finding the ground-state of an Ising spin-system represents one of the most known examples in physics \cite{Barahona_1982, Bachas_1984, Nishimori2001, Ising-NP, Anderson}. 

Interestingly, Ising dynamics have been observed in numerous quantum \cite{Johnson2011,Boixo2014} and classical systems such as random lasers \cite{rand-laser-1, rand-laser-2}, superconducting networks \cite{Harris2018,Hamerly2019}, polariton condensates \cite{Berloff2017,Kalinin2020}, nonlinear wave propagation in disordered media \cite{non-lin-wave-prop}, and degenerate optical parametric oscillators \cite{Yamamoto-1, Yamamoto-2, Haribara_2016,McMahon2016,Inagaki2016,Bohm2019,PhysRevApplied.13.054059,Honjo2021,davidson:2013,yamamoto:2020,strinati:2021, davidson:2020-1,davidson:2021-2}.
Therefore, all aforementioned systems represent suitable platforms to implement NP-problem solvers. For example, coherent Ising-machines using degenerate optical parametric oscillators find approximate solutions for the ground-state of spin-systems\textcolor{black}{, thanks to their inherent non-linearity,} with great control on their couplings. Although optical parametric oscillators can now implement thousands of spins \cite{Honjo2021}, their scalability remains limited by electronic circuits. 

\textcolor{black}{Recently, a second class of optical simulators based on linear optics was demonstrated by exploiting spatial light modulation.} \cite{Pierangeli:19,Pierangeli:20-3,Pierangeli:20, PierangeliMarcucci2020, Leonetti:20,fang2020, fang2021}
In these Ising-machines—where the spins and their interactions are described by phase and amplitude modulation \cite{Pierangeli:19,Pierangeli:20-3,Pierangeli:20, PierangeliMarcucci2020,Leonetti:20} or by a properly engineered gauge-transformation on the optical wavefront \cite{fang2020, fang2021}—the ground-state is found by optimising the detected intensity \textcolor{black}{via a recurrent electro-optical feedback.}
\textcolor{black}{The advantage of this approach relies on the instantaneous calculation of the energy, as the necessary matrix products are encoded in light propagation and are, therefore, independent of the number of spins.}

However, photonic Ising-machines \textcolor{black}{based on wavefront-shaping} using free-space propagation are limited to the implementation of a specific class of hamiltonians, known as Mattis models \cite{mattis}, where the couplings between spins are correlated and an exact solution of the energy ground-state exists. 
Recently, \textcolor{black}{a more general class of hamiltonians, known as Sherrington-Kirkpatrick (SK) spin-glasses \cite{Mezard1986, Parisi1988}, where the couplings are all-to-all and random} was simulated by combining wavefront-shaping and light propagation inside a random medium \cite{Pierangeli:20, Leonetti:20}. Although this optical simulator is scalable \textcolor{black}{and can outperform conventional hardware for a large number of spins} \cite{Pierangeli:20}, its applicability was limited to random all-to-all couplings \textcolor{black}{with zero average}, as determined by the transmission matrix of a random medium.

\begin{figure*}[t]
    \centering
    \includegraphics[width=\linewidth]{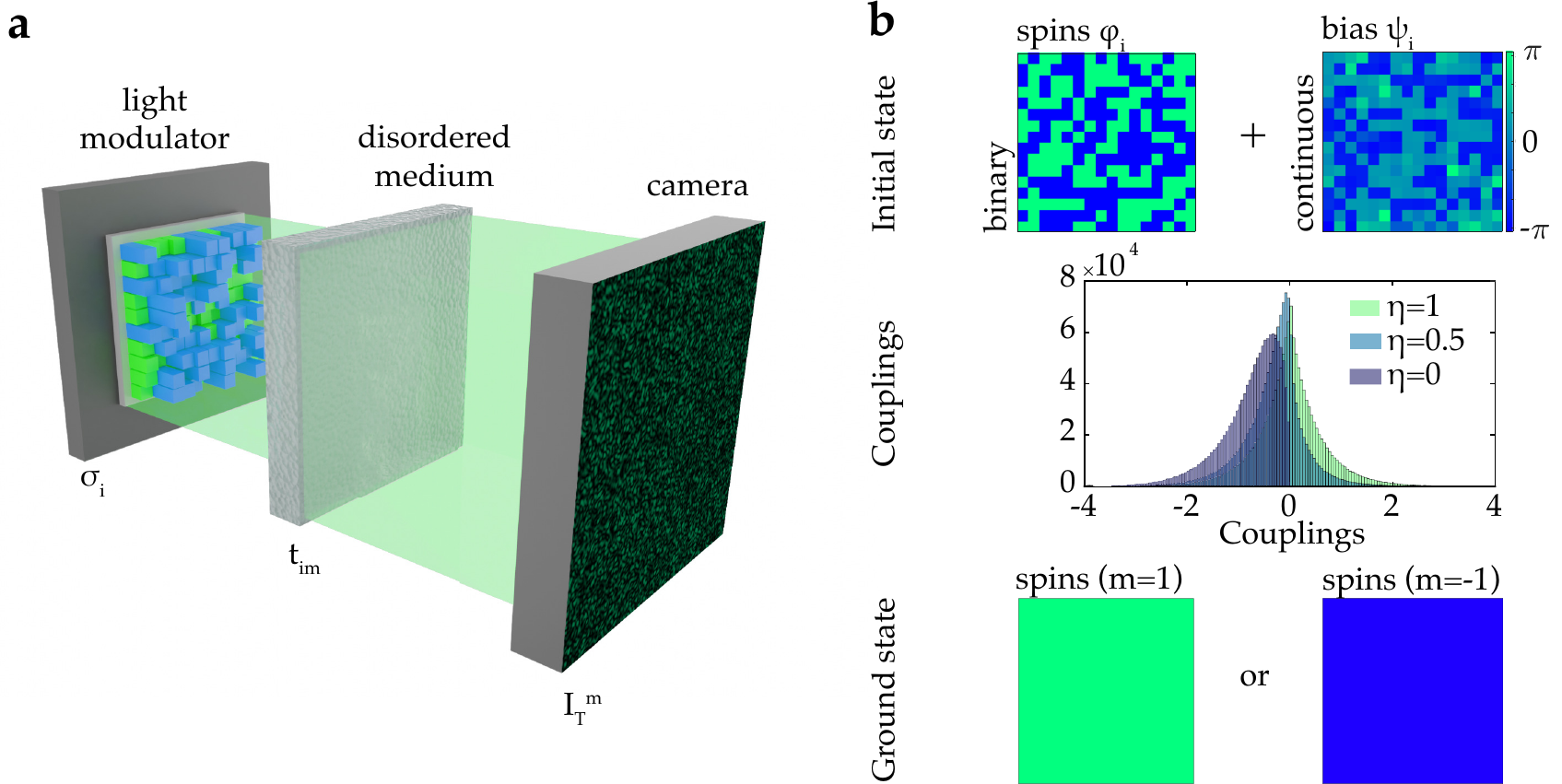}
    \caption{\textbf{Concept of a tuneable optical Ising simulator.} a) Schematics of the Ising-machine based on multiple scattering. The spins are encoded in a binary phase on the spatial light modulator. The ground-state of the Ising Hamiltonian corresponds to the optimised binary phase mask maximising the transmitted intensity recorded by a camera. The all-to-all, random couplings of the spins are induced by multiple light scattering during the propagation in the disordered medium—described by the transmission matrix ($T$). b) Measuring $T$ determines the correct bias ($\psi_i$) to add to the initial phase mask to tune the distribution of couplings. In the case of fully-biased couplings ($\eta=0$), the final state corresponds to a fully-magnetised degenerate ground-state.}
    \label{fig1}
\end{figure*}

In this article, we present various strategies to tune the couplings of a spin-glass Ising simulator based on multiple light scattering.
We experimentally demonstrate the ability to control the evolution of a system in a deterministic fashion by exploiting the knowledge of the transmission matrix of the scattering medium \cite{2011-Popoff}. 
This allows us to observe the phase transition from a disordered to a fully-magnetised ground-state.
Moreover, our work showcases the possibility of modifying the simulated Ising Hamiltonian by both introducing an external magnetic-field term and by controlling the number of degenerate ground-states.

\section{Results} \label{sec:results}

A system of $N$ spins is described by the following Hamiltonian \cite{Mezard1986, Parisi1988}: 
\begin{equation} \label{eq_H}
    H = -\sum_{i,j=1}^N J_{ij} \sigma_i \sigma_j
\end{equation}
where $\sigma_{\{i,j\}}$ and $J_{ij}$ are the spins and their couplings, respectively. 
For all-to-all random couplings, the $J_{ij}$ are drawn from an i.i.d distribution, and finding the ground-state of \autoref{eq_H}—which is referred to as Sherrington-Kirkpatrick model \cite{Nishimori2001}—represents an NP-hard problem.
This specific problem can be mapped into light-propagation in disordered media, where the spins are encoded on a set of input modes or pixels with a binary phase state of $0$ and $\pi$, corresponding to $\pm 1$ spin states, respectively \cite{Pierangeli:20, Leonetti:20}. 
This equivalence can be made explicit by writing the transmitted intensity after a multiply-scattering medium as \cite{Pierangeli:20}:
\begin{equation}
    I_T = \sum_{i,j=1}^N J_{ij} \sigma_i \sigma_j = -H
    \label{eq_I-H}
\end{equation}
with $J_{ij} = -\sum_m^M \Re{\overline{t_{im}} t_{jm}}$, where $m$ runs on the output modes and $t_{im}$ and $t_{jm}$ are transmission matrix ($T$) elements.
\textcolor{black}{$T$ is linked to the input modes on the SLM ($E_{in}$) and the output modes on the camera ($E_{out}$) via $E_{out}=T E_{in}$ \cite{2011-Popoff}.}

\autoref{eq_I-H} shows that maximising the light intensity over the selected output modes allows to retrieve the ground-state of the spin-system (details on the numerical framework used can be found in \refApp{app_num}) \cite{Pierangeli:20}.
In particular, as the distribution of couplings is Gaussian and centred around zero, the ground-state corresponds to an ensemble of randomly oriented spins (mean magnetisation $m=\sum_{i}^N \sigma_i=0$). 
\begin{figure*}[t]
    \centering
    \includegraphics[width=\linewidth]{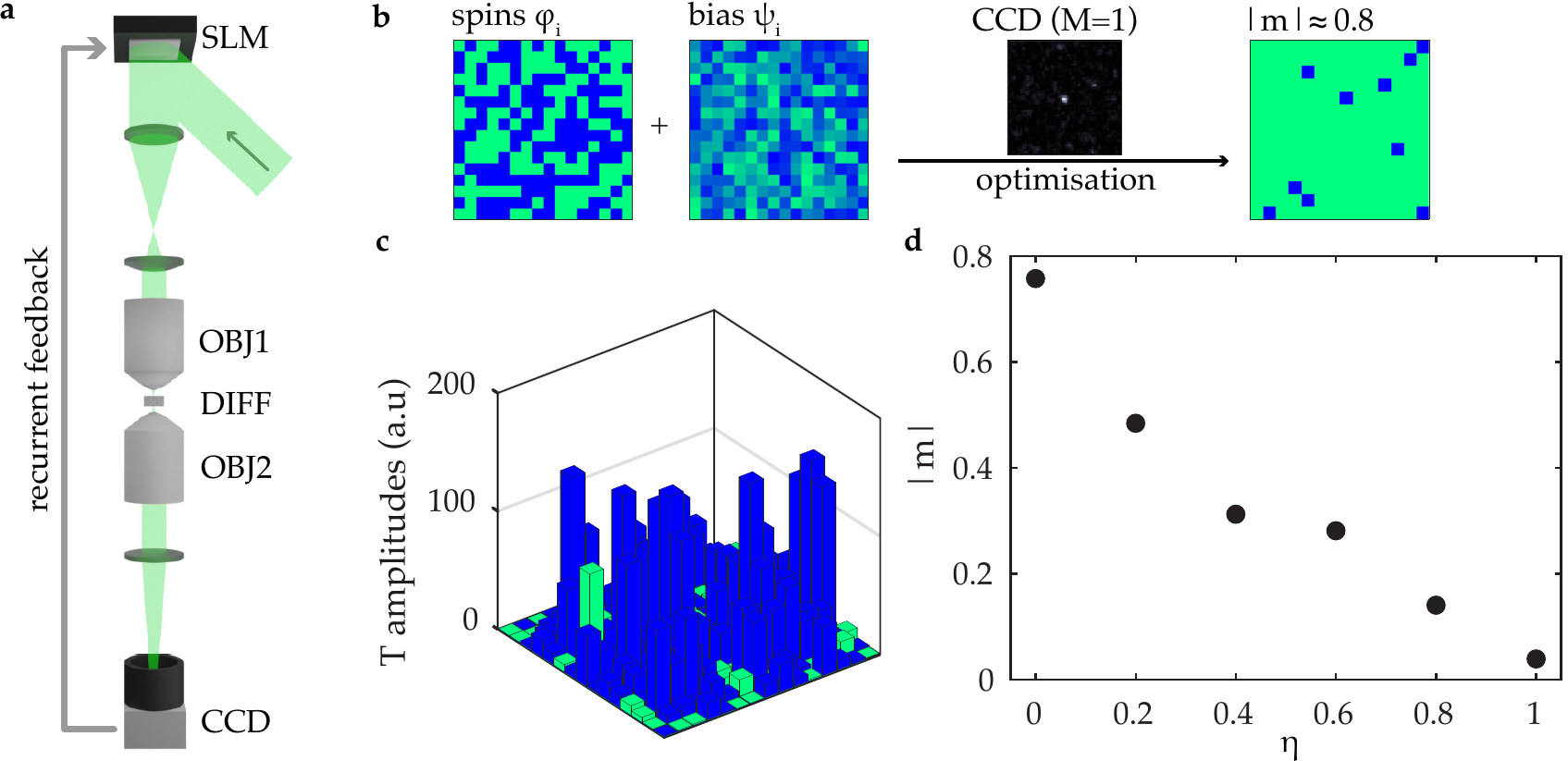}
    \caption{\textbf{Tuneable experimental spin-glass simulator (M=1).} a) Schematics of the experimental setup. Recurrent feedback from the measured intensity on a CCD updates the SLM configuration to reach the SG ground-state. b) Adding a bias to the spin-system ($N=256$ spins) results in a more magnetised ground-state. Inset: intensity focus on the CCD plane as the ground-state is reached. c) The relatively low magnetisation compared to the simulation value can be understood in terms of the finite detection sensibility of the CCD. The spins which are not aligned with the dominant orientation (green (light gray) bars) are indeed those whose T-amplitudes, and therefore intensity-contribution, are the lowest. d) The degree of magnetisation of the ground-state can be readily tuned by adding an artificial noise—a mask of random phases of amplitude $\eta$—to the bias.}
    \label{fig2}
\end{figure*}

As shown in \autoref{fig1}\textcolor{blue}{b}, the $J_{ij}$ distribution can be modified by acting on the initial phases on the SLM. 
Indeed, such an operation leads to the following effective transmission matrix ($T'$): 
\begin{equation}
    T' = T \times \mathrm{diag}(e^{i \psi_i})
    \label{eq_TM'}
\end{equation}
where $\psi_i$ represents the bias associated to each spin. 
\textcolor{black}{\autoref{eq_TM'} rewritten in terms of matrix coefficients becomes:}
\begin{equation}
    t_{im}^{'} = t_{im} e^{i \psi_i}
    \label{eq_TM'_coeff}
\end{equation}
therefore, the resulting couplings—for simplicity still indicated as $J_{ij}$—where $J_{ij} = - \sum_m^M \Re{\overline{t_{im}^{'}} t_{jm}^{'}}$, can be modified by tailoring the bias.

An efficient way to tune the couplings is to exploit the knowledge of $T$. In particular, for a single output mode ($M=1$), using as $\psi_i$ the phase-conjugation of the corresponding single-row $T$ ($\psi_i = -\mathrm{arg}(t_i)$) gives rise to a $J_{ij}$ distribution of only negative values. This results from the fact that after phase-conjugation $T'$ is real-valued. \autoref{fig1}\textcolor{blue}{b} shows also that, consistently with having a negative average coupling $\langle J_{ij} \rangle$, the ground-state of the spins network is not anymore random but it exhibits a fully-magnetised state with degenerate orientations ($m = \pm 1$). 

This result can also be understood in the framework of wavefront shaping \cite{RevModPhys.89.015005}. The introduced bias is the phase mask that optimises the detected intensity, of which the binary spin mask represents an undesired perturbation. Therefore, the optimal intensity is retrieved when the spin mask is reduced to a uniform phase—which in terms of spins corresponds to a fully-magnetised ground-state.

The shift of the $J_{ij}$ can be further controlled by adding a noise mask (mask of random phases) to the phase-conjugation solution. This noise mask is modulated by an amplitude parameter ($\eta$) controlling the shift of the couplings, as shown in \autoref{fig1}\textcolor{blue}{b}. A derivation of the effect of noise on the couplings is presented in \refApp{app_ana}.

We tested experimentally our approach for controlling the couplings of the spin simulator (\autoref{fig2}\textcolor{blue}{a} and \refApp{app_exp}).
A laser (Coherent Sapphire SF 532, $\lambda = 532 nm$) is directed onto a reflective phase-only, liquid-crystal SLM (Meadowlark Optics HSP192-532, $1920 \times 1152$ pixels) divided into N macro-pixels (spins). The modulated light is projected on the objective back focal-plane (OBJ1, $10 \times$, $\mathrm{NA} = 0.1$) and focused on a scattering medium made of Teflon (DIFF) with $0.5 mm$ thickness. The scattered light is collected by a second objective (OBJ2, $20 \times$, $\mathrm{NA} = 0.4$) and the transmitted intensity is detected by a CCD camera (Basler acA2040-55$\mu$m, $2048 \times 1536$ pixels).
The spins and the desired bias are encoded by a spatial light modulator (SLM) in a phase pattern whose binary part is sequentially updated until the ground-state is reached. 
\textcolor{black}{Note that for the optimisation
any algorithm can be used, i.e., the setup is algorithm agnostic, as the advantage of the presented simulators resides in the parallel measurements of the energy.\cite{Pierangeli:20}}

\autoref{fig2}\textcolor{blue}{b} shows the results for a single output mode ($M=1$) on the detection camera (CCD). The ground-state agrees with the numerical predictions of \autoref{fig1}\textcolor{blue}{b}, showing a high magnetisation. \autoref{fig1}\textcolor{blue}{b} shows a typical ground-state averaged over thermal fluctuations \cite{Pierangeli:20-3}—see \autoref{figS1}\textcolor{blue}{a} for single realisations. 

The difference in numerical (cf. \autoref{fig1}\textcolor{blue}{b}, $\abs{m}=1$) and experimental results (cf. \autoref{fig2}\textcolor{blue}{b}, $\abs{m} \simeq 0.8$) can be explained by the contributions that each SLM pixel bears. 
Indeed, as shown in \autoref{fig2}\textcolor{blue}{c}, the spins not aligned with the dominant orientation correspond to SLM pixels with very small T amplitudes—i.e., they marginally contribute to the output intensity and therefore cannot be detected over the optimisation. 
\textcolor{black}{The limited sensitivity of the detection camera introduces an experimental noise term, which allows the spins with small T amplitudes to have any possible orientation. This is equivalent to state that the experimental Ising simulator has an effective, finite temperature \cite{Pierangeli:20}.}
\autoref{fig2}\textcolor{blue}{c} is corroborated by \autoref{figS1}\textcolor{blue}{b}, showing that the light intensity retrieved after optimisation is close to that obtained from phase-conjugation even if the magnetisation does not reach its theoretical maximum. 

Moreover, building upon $T$ it is possible to drive the optical spin-system over a phase transition.
By adding an artificial noise—controlled by the amplitude $\eta$ (cf. \refApp{app_ana})—to the phase-conjugated mask enables to control the degree of magnetisation reached in the ground-state (cf. \autoref{fig2}\textcolor{blue}{d}). This can also be understood in terms of couplings: as predicted in \autoref{fig1}\textcolor{blue}{b}, introducing artificial noise effectively tunes the $J_{ij}$ distribution. 

\begin{figure}[t]
    \centering
    \includegraphics[width=\linewidth]{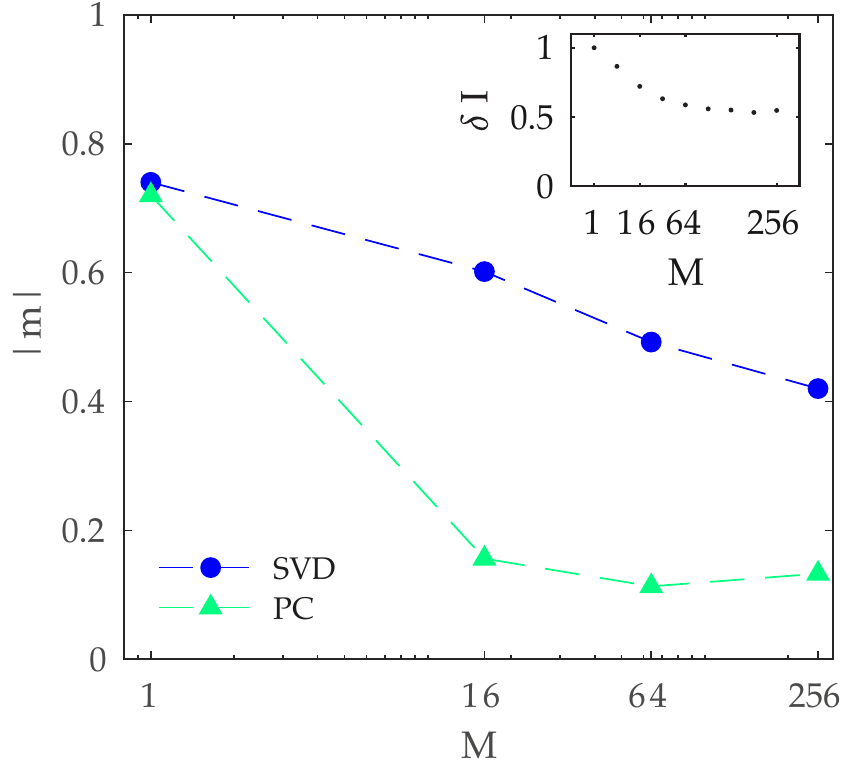}
    \caption{\textbf{Tuneable experimental spin-glass simulator ($M>1$).} Magnetisation as a function of the number of output modes $M$ on the CCD. Using the first singular value (SVD) of the $T$ yields a magnetisation notably larger than with phase-conjugation (PC). 
    This can be understood by comparing the relative focus intensity ($\delta I$) of the two methods: numerical simulations show that the SVD outperforms the PC when increasing $M$, with a two-fold enhancement for a square $T$ ($N=M=256$).
    The top inset displays the intensity distribution on the CCD for the two approaches—for a square T the SVD yields a less homogeneous intensity distribution compared to PC.}
    \label{fig3}
\end{figure}

\autoref{fig2} summarises the case of $M=1$, which simplifies \autoref{eq_H} to the class of Mattis Hamiltonian where the $J_{ij}$ are correlated \cite{mattis,Pierangeli:20,Leonetti:20}. However, such systems are described by an Ising simulator where light propagates in free space \cite{Pierangeli:19,Pierangeli:20-3,fang2020,fang2021}.
To take advantage of the all-to-all \textcolor{black}{random} couplings introduced by the scattering medium is necessary to focus on a higher number of output modes ($M>1$).
Maximising the optical intensity on an area of multiple pixels represents a more complex wavefront shaping task as $T$ has now a size of $M \times N$ and the number of accessible input is N.
The corresponding spin couplings satisfy $\mathrm{rank}(J_{ij})=N$, which implies a NP optimisation problem for the ground-state search \cite{Pierangeli:20}.

\begin{figure*}[t]
    \centering
    \includegraphics[width=\linewidth]{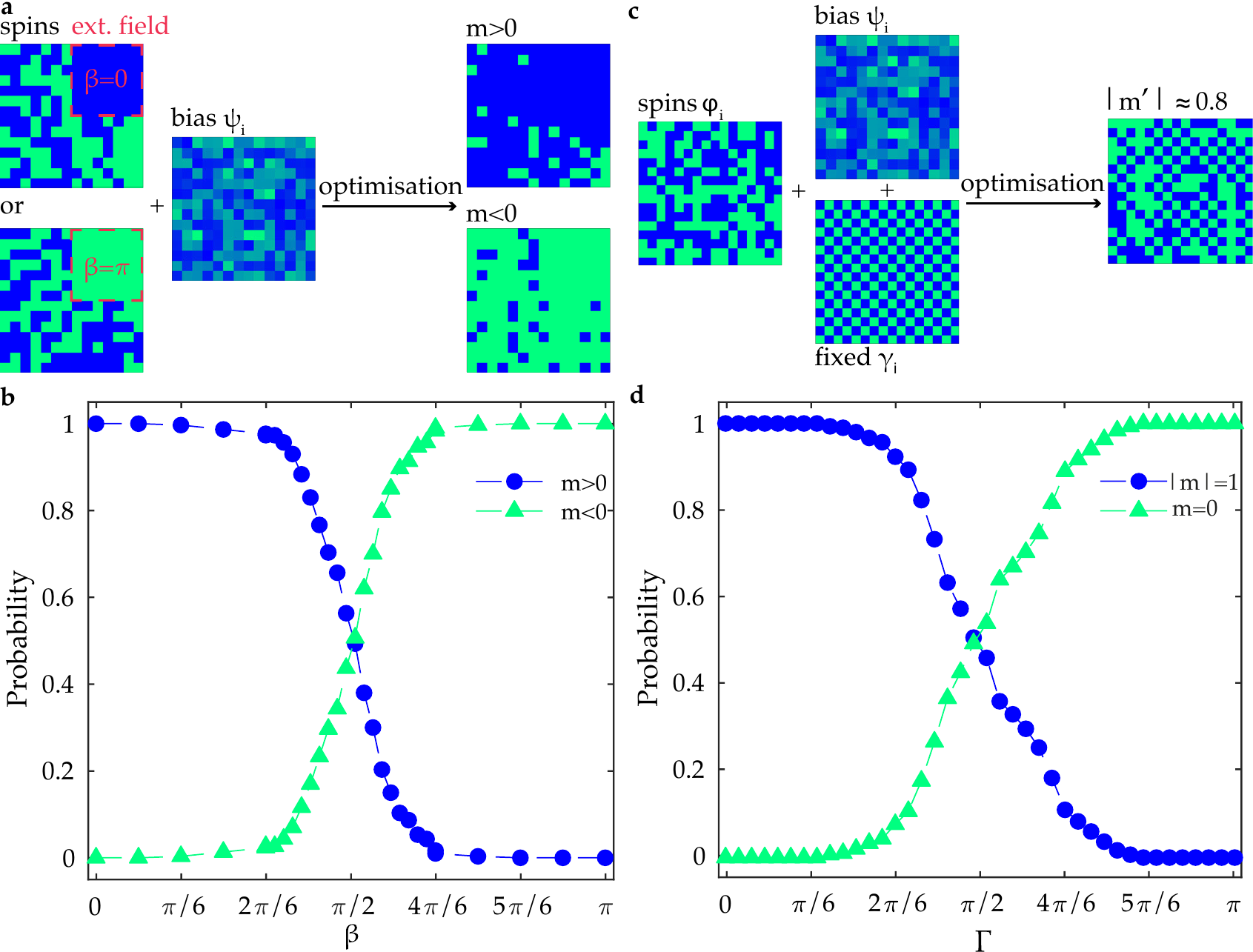}
    \caption{\textbf{Modifying the hamiltonian of an Ising simulator to control the ground state of a spin system.} a) A constant external magnetic-field is implemented by fixing a section of the SLM. The angle associated to the fixed area ($\beta$) determines the final state of the spins system. b) Tuning $\beta$ allows transitioning between the two magnetised-states with opposite orientations. \textcolor{black}{c) Adding a binary pattern ($\gamma_i$) to the bias is possible to determine the spatial distribution of the spins in the ground state. In particular, by using a chequerboard with binary values of $0$ and $\pi$ results in an antiferromagnetic-like ground state — whose fidelity to the bias was quantified in terms of staggered magnetisation ($\vert m^{'} \vert$). d) Changing the value of $\Gamma$) in the chequerboard part controls the probability of the obtaining a ferromagnetic ($\vert m\vert=1$) or an antiferromagnetic-like ($m=0$) ground state. Remarkably, when $\Gamma = \pi /2$, the Ising Hamiltonian presents three ground states — with a degenerate probability for the ferro- and antiferro- magnetic states.}}
    \label{fig4}
\end{figure*}

In the following, we exploited two approaches to choose the optimal bias for $M>1$ modes: i) doing the phase-conjugation (PC) of the sum of the T rows; ii) performing the singular value decomposition (SVD) of $T$  and use the argument of the first right singular vector—i.e., the vector associated to the largest singular value \cite{PhysRevB.23.6851,genack2015,PhysRevLett.117.086803}.  
More details on i) and ii) in \refApp{app_exp}.
The SVD results are more effective in inducing a magnetisation as the T size increases (cf. \autoref{fig3} and \textcolor{blue}{Figure S3a} for experimental and numerical results, respectively).  

Indeed, the SVD mask leads to a higher focus enhancement than the PC one, with a twofold enhancement for a square $T$ ($N=M=256$), and it produces a more marked shift of the $J_{ij}$ distribution (cf. inset of \autoref{fig3} and \autoref{figS2}, respectively).

A decrease of the magnetisation in function of M can again be understood in terms of T amplitudes. \autoref{figS3}\textcolor{blue}{b} shows that for $M=256$ a correlation between the alignment of the spins and their T amplitudes is observed. However, compared to $M=1$ (c.f. \autoref{fig2}), the T amplitudes are more homogeneously distributed and decrease in value—meaning that more spins have a marginal, experimentally undetectable, contribution to the intensity.

In \autoref{fig2} and \autoref{fig3} we demonstrated that, by measuring $T$, a magnetised ground-state can be promoted in a spin-system with all-to-all couplings. 
Importantly, the presented approach can also be used to generalise \autoref{eq_H} accounting for an external magnetic-field. 
In particular, as derived in \refApp{app_ana}, by selecting a given portion of the SLM which will stay fixed to a specific phase ($\beta$) defined before the optimisation, i.e., by defining m SLM modes with phases  $b_1, b_2, ...b_N = b$, \autoref{eq_I-H} becomes:
\begin{equation} \label{eq_H_extField}
	I_T = \sum_{i,j=1}^N J_{ij} \sigma_i \sigma_j + B \sum_i \sigma_i
\end{equation}
where $B \propto \Re{e^{i \beta}}$ represents the added external field, defined from a fixed phase $\beta \in [0, 2 \pi)$. 

This term represents a constant external magnetic-field, driving the spins towards a specific sign of magnetisation. In the two limiting cases $\beta = 0, \pi$ the machine is expected to settle a non-degenerate ground-state that depends on the value of $\beta$ (cf. \autoref{fig4}\textcolor{blue}{a}). Note that the bias is applied also onto the area of the SLM dedicated to the external field to effectively drive the evolution of the spins.
Moreover, $\beta$ tunes the intensity of the magnetic-field and therefore the transition between the two degenerate spins' orientations (cf. \autoref{fig4}\textcolor{blue}{b}). The sharpness of which is defined by the ratio between the number of SLM pixels allocated to the spins and those describing the magnetic-field (see \autoref{figS4}).

Remarkably, the ground-state found by the spin-glass simulator can be further modified by adding a tailored, binary shift to the T bias ($\gamma$). 
By doing so, the system will converge to the defined shift pattern (cf. \autoref{fig4}\textcolor{blue}{c}). \autoref{fig4}\textcolor{blue}{c} shows that a ground-state resembling an antiferromagnetic system is obtained by defining $\gamma_k=(1-e^{i\pi k})\Gamma$, with $\Gamma \in [0, 2 \pi]$. The fidelity of the ground-state—quantified in terms of staggered magnetisation ($m^{'}$) \cite{stagg-mag}—is close to what was observed in \autoref{fig1}\textcolor{blue}{b}. Moreover, \autoref{fig4}\textcolor{blue}{d} shows that by varying the angle associated with the additional shift mask ($\Gamma$), the energy landscape of the system can be affected. Without any additional shift, the ground-state is fully-magnetised—as already shown in \autoref{fig1}\textcolor{blue}{b}. When $\Gamma = \frac{\pi}{2}$, the system has an additional equiprobable ground-state which becomes dominant when $\Gamma = \pi$. The form this shift and corresponding ground-state take can be extended to any arbitrary mask \autoref{figS5}.

\section{Conclusion}
In conclusion, we reported a novel approach to tailor the Hamiltonian of all-to-all coupled spin-systems implemented on a spatial photonic Ising-machine. In contrast to previous works, our method controls the couplings by shifting their distribution. 
Such tuneability was obtained by exploiting the complex coefficients of the transmission matrix and tuning the number of output modes.
In particular, knowing the transmission matrix allows to map a Sherrington-Kirkpatrick spin-glass model with tuneable couplings and an external field term through a co-propagating reference on the input wavefront.
Finally, we demonstrated that greater control over the energy landscape—and hence the possible ground-states and their likelihood—can be obtained by introducing a tailored binary phase shift.

The accuracy of the Ising-machine, quantified in terms of magnetisation and transmission matrix amplitudes, can be improved by reducing the experimental noise of the setup.
However, to target specific couplings a different approach is required. In particular, we foresee two possible routes: 
i) develop algorithms that takes advantage of the huge number of pixels of both the SLM and the camera to choose or group them to re-define a spin-system with the target couplings; 
ii) replace the random medium with an engineered system that maps the desired couplings.
Although computationally and experimentally challenging, these strategies may unlock the ability to have a fully-tuneable spin-glass Hamiltonian.
\textcolor{black}{Moreover, the performance of the Ising-machine could be further improved by replacing the electro-optical feedback in the ground-state search with an all-optical, wavefront-shaping approach \cite{Nixon2013}}

\begin{acknowledgments}
We thank Dr. R. Pandya and Dr. J. Dong for useful discussions. G.J. and L.D. contributed equally to this work. G.J., L.D., D.P., M.R., C.C., and S.G. designed the project.  L.D. and G.J. carried out experiments, numerical simulations and data analysis. G.J. and L.D. wrote the paper with contributions from all the authors.  This project was funded by the European Research Council under the grant agreement No. 724473 (SMARTIES). D.P. and C.C. acknowledge funding from SAPIExcellence BE-FOR-ERC 2020 Grant, the QuantERA European Research Area Network (ERA-NET) Cofund [Grant No. 731473, the “Quantum Information Processing with Complex Media” (QuompleX) project], the Progetti di Ricerca di Interesse Nazionale (PRIN) 2017 program (Photonic Extreme Learning Machine -PELM project). M.R. acknowledges funding from Science Committee of Armenia under the grant agreements No. 20TTSG-1C008 and No. 21APP-1C016. 
\end{acknowledgments}

\appendix
\renewcommand{\thesection}{\Alph{section}}
\section{Experimental details}
\label{app_exp}
\subsection*{Transmission matrix calculation and ground-state search}
The transmission matrix of the scattering medium was estimated as in \cite{2011-Popoff}. In detail, each row of the TM can be reconstructed by monitoring how the intensity on a given CCD pixel changes when a phase modulation is applied to the input patterns on the SLM.  The TM is sensitive to translations and rotations of the scattering medium as well as to the input and detection hard- ware. In this work, we define the stability as a variation within 10\% of it original value. The time is long enough to run our experiments but for larger systems one would need more stable architectures.

The ground-state search is conducted sequentially by means of the recurrent digital feedback. Computation starts from a random configuration of N binary macro-pixels (spins) on the SLM. The measured intensity distribution determines the feedback signal.  At iteration, a arbitrary batch of spins is randomly flipped if it increases the intensity at a chosen output mode. The batch size decreases over the optimisation procedure, starting from $12\%$ of the pixels to a single pixel for the last $\sim 600$ iterations. Note that our simulator is agnostic to the optimisation algorithm used.

\subsection*{Focusing on multiple output modes: phase conjugation and singular value decomposition}
\label{SI_PCvsSVD}
The ideal option to perfectly focus through a medium would be to be able to invert the transmission matrix. But this is extremely hard to do in practice we only use the phase-conjugated matrix. Each element of this matrix is defined as a complex element that has a magnitude of $1$ and an argument equal to the opposite of that specific element in the initial transmission matrix, i.e. if $\theta$ is the argument of an element in the 
TM then $-\theta$ is the argument of the corresponding conjugated matrix.

For several output modes, one first needs to sum over the rows of the initial TM and then extract the arguments. Indeed, we only control $N$ degrees of freedom thanks to the SLM but for several output modes the TM is $N \times m$. Once this is done one can actually use the opposite angle to counteract the disorder medium. The larger the matrix the less easy it is to actually get a proper focus.\\ 

The second method is based on the singular value decomposition (SVD). It is defined as follows: any given matrix $A \in \mathbb{C}$ of dimensions $m \times n$ can be factorised as $U \Sigma V^{\dagger} = A$, where the columns of $U$ (resp $V$) represents the left (resp right) singular vectors and the diagonal matrix $\Sigma$ represents the singular values ranked by descending order — i.e., the most representative vectors of A are the first ones of each matrix. In this particular nomenclature, the matrix $V$ (resp $U$) encodes for the output (resp input) modes. One can therefore use the opposite of the arguments of the first right singular vector (V here) to act as a phase conjugation substitute. Indeed, the \emph{first} singular vector is the one with the biggest singular value and therefore is the  one that represents the most the actual matrix.

\section{Numerical details}
\label{app_num}
The numerical model used in this work is a generalisation of what described in \cite{Pierangeli:20}. The optical SG is numerically simulated by forming N pixel blocks (SLM plane). The initial optical field has a constant amplitude, and its phase is a random configuration of N binary phases, $\phi_i=0, \pi$. A transmission matrix T with random complex numbers is generated. At each iteration, a randomly selected single spin is flipped. The input phase updates if the output total intensity increased after the linear propagation of the field. The bias in the numerical framework is calculated as in the experiment by starting from the knowledge of T.

Numerical evaluation of $I_T$ corresponds to a measurement with a detector in a noiseless system. In general, within this scheme, $\sim 10N$ iterations are sufficient for a good convergence, i.e., when focus intensity reaches a plateau.  All codes are implemented in MATLAB on an Intel processor with 14 cores running at 3.7 GHz and supported by 64 GB ram.
\\

\section{Details of the analytical calculations}

\label{app_ana}
\subsection*{Noise bias}
\label{SI_noise_couplings}
Additionally to the context of having the spins superimposed with the SVD retrieved angles to compensate for the medium, one can define a random mask superimposed on top of this which reads as: 
$e^{i \theta \eta}$, where $\theta$ is a random value in $[0,2\pi]$ and $\eta \in [0,1]$ is a global parameter which defines how strong this noise is compared to the SVD angles ($\eta=1$ meaning the noise is a strong and therefore the situation is equivalent to not having retrieved any angles at all).
Consequently, the intensity of one output mode $m$ reads as:
\begin{align*}
    I_T =& |\sum_i t_{im} E_i|^2\\
    =& \left(\sum_j \overline{t_{jm} E_j}\right)\left(\sum_i t_{im} E_i\right)\\
    =& \sum_{i,j} t_{im} \overline{t_{jm}} E_i \overline{E_j}\\
    =& \frac{1}{2} \left(2 \sum_{i,j} t_{im} \overline{t_{jm}} E_i \overline{E_j}\right)
\end{align*}
where $t_{im} = A_i e^{-i \Tilde{\psi_i}}$ are the TM coefficients and $E_i = e^{i \phi_i} e^{i \psi_i} e^{i\eta \theta_i}$.
The intensity and resulting couplings can therefore be rewritten as:
\begin{align}
    I_T =& \sum{i,j} \widetilde{J_{ij}}\sigma_i \sigma_j\\
    \widetilde{J_{ij}} =& -\Re{t_{im} e^{i \psi_i} e^{i\eta \theta_i} \overline{t_{jm}} e^{-i \psi_j} e^{-i\eta \theta_j}}
\end{align}

\subsection*{External magnetic field}
\label{app_expF}
To derive the effective external field, one needs to separate the SLM into two parts, one portion encodes for the spins — optimised by the algorithm — and one portion encodes for the external field — fixed during optimisation. Therefore, considering the case where half the SLM is allocated to the spins $\sigma_i=e^{i\phi_i}$ as well as to the external field $h_j=e^{i \beta_j}$, one can divide the usual sum into two terms : 
\begin{align*}
    I_T &= |\sum_i t_{im} E_i|^2\\
    &= \left(\sum_{j=1}^{N/2} \overline{t_{jm} E_j} + \sum_{j=N/2+1}^{N} \overline{t_{jm} E_j}\right)\\
    &\times \left(\sum_{i=1}^{N/2} t_{im} E_i + \sum_{i=N/2+1}^{N} t_{im} E_i \right)\\
    &= \sum_{i,j} \overline{t_{jm}}t_{im} \overline{E_j} E_i + \sum_{i,j} \overline{t_{jm}}t_{im} \overline{E_j} E_i\\
    &+ \sum_{j=1}^{N/2} \sum_{i=N/2+1}^{N} \overline{t_{jm}}t_{im} \overline{E_j} E_i + \sum_{i=1}^{N/2} \sum_{j=N/2+1}^{N} \overline{t_{jm}}t_{im} \overline{E_j} E_i
\end{align*}
where $E_i = e^{\phi_i + \psi_i}\text{ or }e^{\beta_i + \psi_i}$ depending on whether it corresponds to the spin area or the field area, respectively.\\

The first term is the one derived earlier which represents the classical Hamiltonian. The second one is simply a constant term based on the field value which will not change during optimisation — which will be neglected in the coming lines. Acting on the index order we obtain: 
\begin{align*}
    I_T &= \sum_{i,j} J_{ij} \sigma_i \sigma_j  + \sum_{j=1}^{N/2} \sum_{i=N/2+1}^{N} \overline{t_{jm}}t_{im} \overline{E_j} E_i\\
    &+ \sum_{i=1}^{N/2} \sum_{j=N/2+1}^{N} \overline{t_{jm}}t_{im} \overline{E_j} E_i\\
    &= -H(\sigma) + \sum_{i=1}^{N/2} \sum_{j=N/2+1}^{N} (\overline{t_{jm}}t_{im} \overline{E_j} E_i + \overline{t_{im}}t_{jm} \overline{E_i} E_j)\\
    &= -H(\sigma) + \sum_{i=1}^{N/2} \sum_{j=N/2+1}^{N} (\overline{t_{jm}}t_{im} \overline{h_j} e^{-i\psi_j}\sigma_i e^{i \psi_i}\\
    &+ \overline{t_{im}}t_{jm} h_j e^{i\psi_j} \overline{\sigma_i} e^{-i \psi_i})
\end{align*}

By definition, $\sigma_i = e^{i \phi_i}$ and since the spins are binary — $\phi_i = 0,\pi$ — then $\sigma_i = 1,-1$ and therefore $\overline{\sigma_i} = -\sigma_i$. Writing out the TM coefficients as $t_{im} = A_i e^{i\Tilde{\psi}_i}$ and the field as $h_j=e^{i \beta_j}$ one gets :
\begin{align*}
    I_T &= -H(\sigma)\\ &+ \sum_{i=1}^{N/2} \sum_{j=N/2+1}^{N} (A_j e^{i \Tilde{\psi_j}} A_i e^{-i \Tilde{\psi_i}} e^{-i \beta_j} e^{-i\psi_j} e^{i \psi_i} \sigma_i\\
    &+ A_j e^{-i \Tilde{\psi_j}} A_i e^{i \Tilde{\psi_i}} e^{i \beta_j} e^{i\psi_j} e^{-i \psi_i} \overline{\sigma_i})\\
    &= -H(\sigma)\\ &+ \underbrace{\sum_{i=1}^{N/2} \sum_{j=N/2+1}^{N} \Re{(e^{-i (\psi_j - \Tilde{\psi_j})} e^{i (\psi_i - \Tilde{\psi_i})} e^{i \beta_j}} A_i A_j \sigma_i}_{(\ast)}
\end{align*}

Now, the $\psi_{i,j}$ are, by definition, the phase conjugated angles of the TM which implies that in case of an ideal phase conjugation $\psi_i = \Tilde{\psi_i}$. Considering the field is uniform, meaning $\beta_j = \beta$ then this last term can therefore be rewritten as: 

\begin{align*}
    (\ast) &= 2 \sum_i \sum_j \Re{e^{i \beta_j}} A_i A_j \sigma_i\\
    &= 2\Re{e^{i \beta}} \sum_j A_j \sum_i A_i \sigma_i\\
    &= 2\Re{e^{i \beta}} \left(\sum_j A_j\right) \sum_i A_i \sigma_i\\
\end{align*}

We therefore obtain:
\begin{equation*}
    I_T = \sum_{i,j} J_{ij} \sigma_i \sigma_j  + B \sum_i A_i \sigma_i
\end{equation*}
where $B = 2\Re{e^{i \beta}} \sum_j A_j$ describes the effective external field.

\section{Additional figures}
\setcounter{figure}{0}
\renewcommand{\thefigure}{A.\arabic{figure}}
\begin{figure*}[t]
    \centering
    \includegraphics[width=\linewidth]{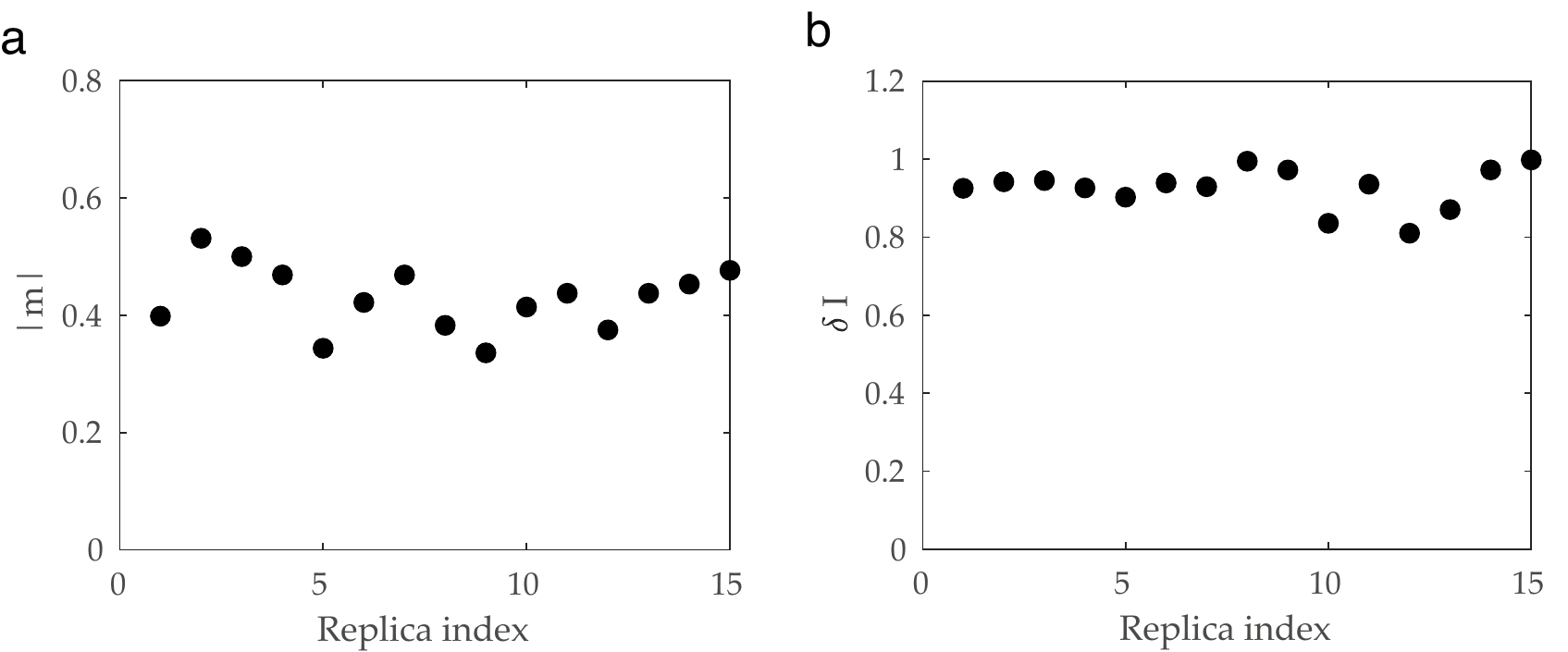}
    \caption{a) Magnetisation values of different experimental realisations. Averaging the spins optimal mask allows to account for thermal fluctuations and obtain a reliable ground-state (as shown in Fig1b, main text) to compare to the corresponding theoretical zero-temperature solution. A noise-driven annealing approach does not improve the performance of the optimisation. b) Comparison between the focus intensity obtained via optimisation and from phase-conjugation of the TM for M=1. The relative focus intensity ($\delta I$) is close to unity, showing that the optimisation correctly reaches the highest focus intensity.}
    \label{figS1}
\end{figure*}

\begin{figure*}[t]
    \centering
    \includegraphics[width=\linewidth]{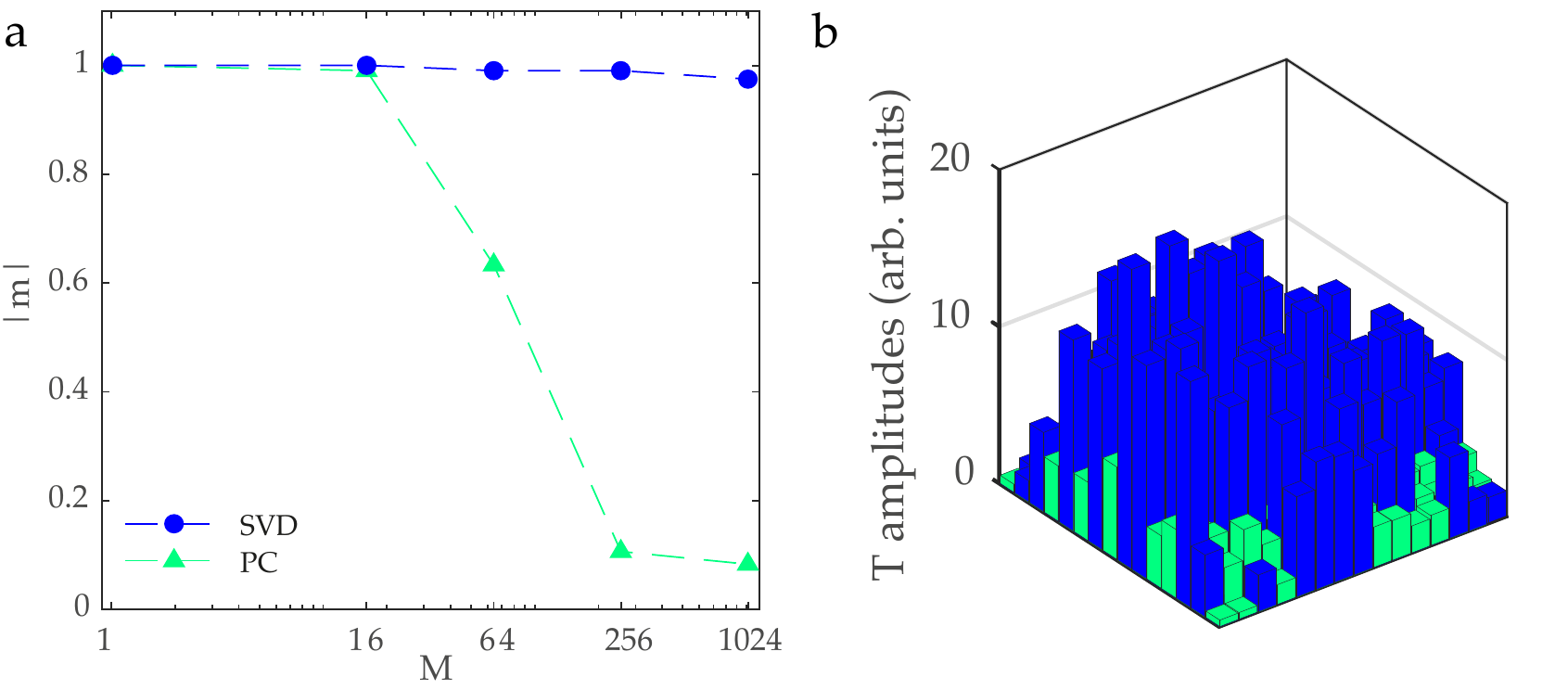}
    \caption{a) Simulation of the expected magnetisation in function of the number of output modes for both SVD and PC biases. b) TM amplitudes associated to the SLM spins when considering M=256. As in \autoref{fig2}, a correlation between the alignment of the spins and their TM amplitudes is observed. Compared to M=1 the TM amplitudes are more homogeneously distributed and significantly decrease in value, leading to a lower experimental magnetisation.}
    \label{figS4}
\end{figure*}

\begin{figure}[t]
    \centering
    \includegraphics[width=\linewidth]{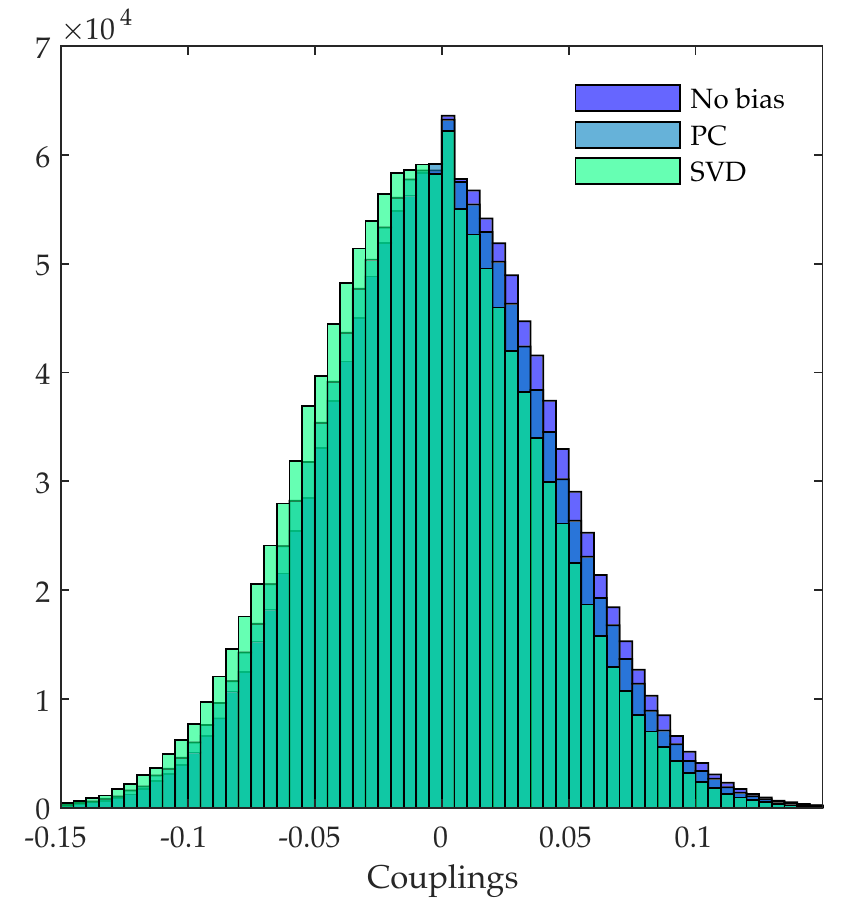}
    \caption{Couplings distributions for systems with M=N=256. In the case of SVD, the distribution results more shifted than when using PC. This is in agreement with what observed in \autoref{fig3} of the main text in terms of magnetisation of the ground state.}
    \label{figS2}
\end{figure}

\begin{figure}[t]
    \centering
    \includegraphics[width=\linewidth]{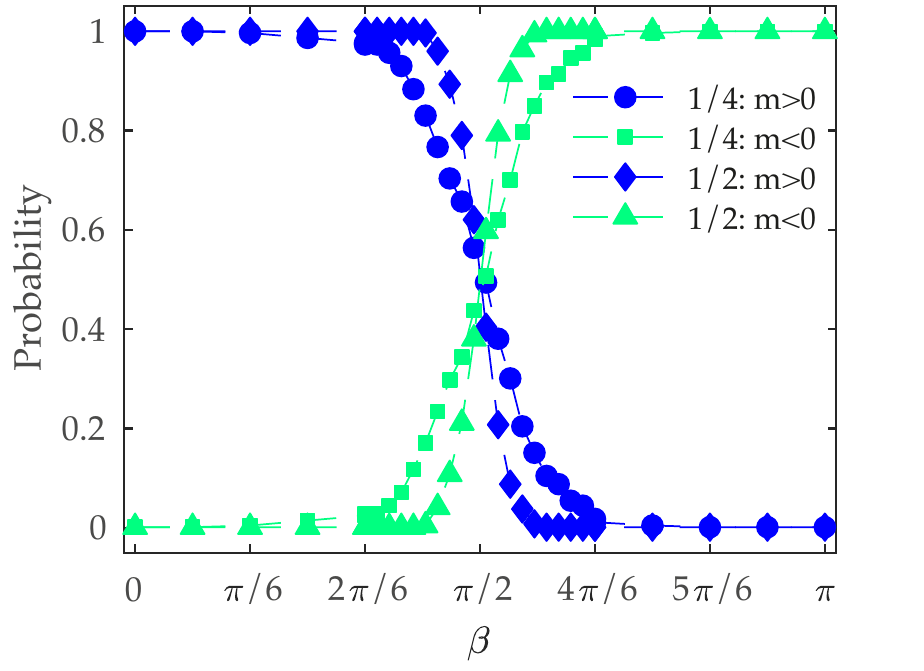}
    \caption{Allocating a larger area of the SLM to the magnetic field results in a sharper phase transition.}
    \label{figS5}
\end{figure}

\begin{figure}[t]
    \centering
    \includegraphics[width=\linewidth]{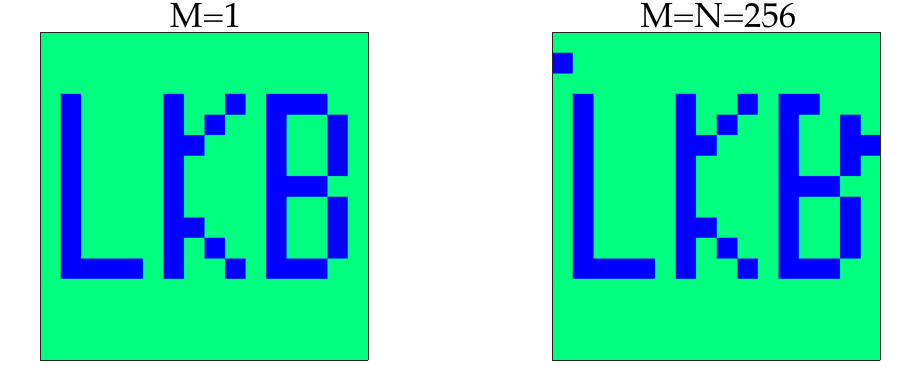}
    \caption{Modifying the energy landscape to control the ground state of an Ising simulator. Example of the spin ground state obtained using an additional binary bias showing the Laboratory Kastler-Brossel initials.}
    \label{figS3}
\end{figure}

\FloatBarrier

\bibliography{main.bib}
\end{document}